\documentclass[graybox, secnum]{svmult} 
\usepackage{lineno}

\usepackage{mathptmx}       
\usepackage{helvet}         
\usepackage{courier}        
\usepackage{type1cm}        
\usepackage{xcolor}
\usepackage{siunitx}
\usepackage{makeidx}         
\usepackage{graphicx}        
\usepackage{subfig}                             
                             
\usepackage{multicol}        
\usepackage[bottom]{footmisc}
\usepackage{hyperref}        
\usepackage{soul}            
\hypersetup{colorlinks=true,urlcolor=blue}
\usepackage[square,numbers]{natbib}
\makeindex             

\begin{document}
\title*{Gamma-ray detector and mission design simulations}
\author{Eric A. Charles and Henrike Fleischhack
  and Clio Sleator}
\institute{Eric A. Charles \at Kavli Institute for Particle Astrophysics and Cosmology, SLAC National Accelerator Laboratory,  Menlo Park, CA 94025, USA \email{echarles@slac.stanford.edu}
\and Henrike Fleischhack \at Catholic University of America, Washington, DC 20064, USA; NASA/GSFC, Greenbelt, MD 20771, USA; CRESST II, Greenbelt, MD 20771, USA \email{fleischhack@cua.edu}
\and Clio Sleator \at U.S. Naval Research Laboratory, Washington DC 20375, USA \email{clio.sleator@nrl.navy.mil}}
%
%
\maketitle
\abstract{Detectors for gamma-ray astronomy are complex: they often comprise multiple sub-systems and utilize new and/or custom-developed detector components and readout electronics. Gamma rays are typically not detected directly: ground-based detectors measure extensive air showers of charged particles initiated by cosmic gamma-rays, and even so-called ``direct detection'' experiments on balloons or satellites usually reconstruct the incoming gamma-ray photons' properties from the secondary particles produced in the detector. At the same time, there are few ``standard candles'' and no feasible terrestrial sources of high-energy and very-high-energy gamma rays that could be used to calibrate the detectors. Simulations of particles interacting in the atmosphere and/or with the instrument are thus ubiquitous in gamma-ray astronomy. These simulations are used in event reconstruction and data analysis, to characterize detector performance, and to optimize detector design. In this chapter, we give an overview of how and why simulations are used in gamma-ray astronomy, as well as their limitations. We discuss extensive air shower simulations, simulations of gamma rays and secondary particles interacting in the detector, and simulations of the readout electronics. We provide examples for software packages that are used for various aspects of simulations in gamma-ray astronomy. Lastly, we describe the performance metrics and instrument response functions that are generated from these simulations, which are critical to instrument design and data analysis. 
}

\section{Keywords} 

gamma rays; simulations; monte-carlo technique; instrument response; air showers; IACTs; balloon-based gamma-ray detection; space-based gamma-ray detection; ground-based gamma-ray detection

\section{Introduction: why we do simulations and how we use them}

In the course of designing and operating gamma-ray telescopes and space missions, many different types of Monte-Carlo simulations are used.   Depending on the type of instrument, these can include: simulations of the interactions between gamma rays and the instrument, simulations of the interactions of the gamma-rays with the atmosphere above the instrument, simulations of interactions of other types of particles with the instrument or the atmosphere, simulations of the fluxes of gamma-rays from celestial sources incident on the instrument or atmosphere, simulations of the fluxes of other types of particles, and simulations of the electronics response to the energy deposited in the active volumes of the detector by incoming particles.

In short, we use these simulations to explore the vastly complicated spaces of source fluxes, particle interaction chains, and detector responses that determine how much of the flux of celestial gamma rays a given instrument will detect, how accurately it will measure interaction properties such as energy and incident direction, and how much the gamma ray sample may be contaminated by other types of particles mis-identified as gamma rays.  More specifically, we use the simulations to characterize the response of a given instrument and instrument configuration to a set of known input fluxes.  We then use these characterized responses in a number of ways.

While designing a telescope or mission, we use characterizations of the instrument performance for ``trade studies'', in which we explore the trade-offs associated with various design choices to arrive at an optimal instrument design.  Going a step further, we can use the same characterizations to arrive at ``sensitivity estimates'', i.e., estimates of the minimum flux detectable given a particular set of observations.  We can then use these sensitivity estimates to explore the science potential of a particular instrument; asking questions such as the number of sources the instrument might detect, or the precision to which it might measure fluxes or source positions.

When operating a telescope or mission, the role of simulations shifts somewhat.  Although some amount of instrument and observing strategy optimization will continue, during operations we primarily use simulations in the model fitting we do as part of the scientific data analysis.  Specifically, most gamma-ray astronomy data analysis is done with ``forward folding'' techniques, wherein we convolve a model of source fluxes with representation of the instrument response obtained from simulations and compare the result to the data obtained by the instrument.  The model fitting procedure then consists of finding the set of model parameters which result in the best fit to the data.

\section{Common aspects of simulations}

\subsection{Astronomical inputs, sources, fluxes, backgrounds}

Astronomical inputs to simulations consist of two components: sources and backgrounds. Astronomical sources can either be point sources, such as Gamma-Ray Bursts (GRBs), neutron stars, etc., or diffuse emission, such as the positron annihilation line from the Galactic center.   When describing sources, in addition to the source extent, the source position relative to the instrument must be specified, since it affects the instrument response: for example, the instrument efficiency is generally higher for a source directly on-axis compared to a source at the edge of the field of view. The source flux and energy spectrum information are related, as the flux of astrophysical sources is energy dependent. The energy spectrum is often given as an empirical function such as a power law, in which case the flux is usually given at a single energy and then extrapolated to all energies. In the case of gamma ray nuclear line sources, the energy spectrum is given as discrete single energies, and potentially the associated line widths, each with a specified flux.

Simulations of the background are required to determine instrument sensitivity estimates, as well as to test certain aspects of analysis methods such as classifying events as source or background. For satellite missions, cosmic rays are the main source of background. Primary cosmic ray particles, predominantly protons, electrons, and heavier ions, can interact directly with the instrument. Additionally, these primary cosmic rays interact in the atmosphere, producing both charged and neutral secondaries. Since the signature of charged particles and neutrons in a detector is often sufficiently distinct from the signature of a gamma ray, charged particle background events are usually straightforward to identify. The gamma rays produced as secondaries, however, are more difficult to separate from source events.

Simulating the interaction of cosmic rays with the instrument is also important to mimic instrument activation. When the instrument is bombarded with primary and secondary cosmic ray particles, nuclear reactions are induced in the instrument material and radioactive isotopes are formed. These radioactive isotopes decay and can emit gamma rays that are difficult to distinguish from source photons.

The spectrum and flux of the cosmic ray background primarily depend on the orbit altitude and inclination in addition to the current solar activity. See \cite{Cumani2019} and \cite{Mizuno2004} for in-depth discussions on cosmic ray background models.

The background environment changes for balloon-borne instruments that float at the upper edge of the atmosphere, and for ground-based instruments that detect the highest energy gamma rays via Cherenkov radiation. At typical altitudes for balloon-borne instruments (30-40~km), the albedo radiation produced by cosmic rays interacting in the Earth's atmosphere is much brighter than at satellite orbital altitudes. Balloon backgrounds are often estimated using models based on data from previous balloon flights (e.g. \cite{Ling1975}).

The dominant backgrounds for ground-based gamma ray detectors are due to air showers induced by (charged) cosmic rays. While simulations are often used for trade studies and to estimate sensitivities, data-driven background estimation methods (e.g. using off-source data) are preferred whenever available due to the uncertainties in both the cosmic-ray fluxes/spectra and the air shower simulations.

Instruments detecting atmospheric Cherenkov emission, i.e., imaging air-Cherenkov Telescopes (IACTs), also suffer from so-called night-sky background (NSB): optical and ultraviolet (UV) light from both astronomical and terrestrial sources. NSB is affected by the phase of the moon, stars in the field of view, and atmospheric conditions (clouds, aerosols). NSB affects IACTs in several ways: first, by adding noise on top of real air shower signals, and second, by increasing the trigger rate due to chance coincidences of noise. In some instruments, bright starlight can also lead to camera pixels being temporarily turned off or removed from the trigger. Purely NSB-induced events can be suppressed by increasing the trigger threshold or applying cuts on the reconstruction quality, at the cost of increasing the effective energy threshold and decreasing the effective area. NSB noise (from measurements or dedicated simulations) is often added to the air shower simulations after the fact, so that the same air shower simulations can be re-used under different NSB conditions.

\subsection{Detector geometries}
In addition to the astronomical inputs, a mass model of the instrument is required for simulations. The mass model describes the type, amount, and location of all of the material that makes up the instrument. In addition to the active detectors, the mass model must also include passive material since gamma rays can interact with the passive material of the instrument. For example, gamma rays that first interact in passive (non-instrumented) material will enter the detectors with a lower energy, which affects the measured spectrum, the event reconstruction, and the general instrument response; it is important to include these effects in simulations. The passive material closest to the detectors should be modeled with as much detail as possible. For a space-based instrument, the spacecraft bus should also be included in the mass model.

Instrument activation can occur when cosmic rays interact in the instrument material and produce radioactive isotopes, which then decay. These decays can produce gamma rays which are then detected by the instrument. Simulating this activation background component requires an accurate mass model, since the half-life of the radioactive isotopes and the energy of the emitted gamma rays are dependent on the specific materials used to build the instrument.

\subsection{Physics Input}

Simulation engines used for gamma-ray astronomy have to take into account a variety of possible interactions with gamma-ray photons and charged particles with the Earth's atmosphere and/or the detector, as described in the following subsections.

\subsubsection{Extensive air showers}

When gamma rays with energies of tens of MeV or more strike the atmosphere, they primarily interact via production of electron-positron pairs in the electric fields of ambient electrons or ambient nuclei. If the newly produced electrons and positrons are energetic enough, they can in turn emit gamma rays via Bremsstrahlung in the same electric fields, which can produce more electron-positron pairs and so on. This process continues until the energy of the electrons and positrons falls below the critical energy of about \SI{90}{\MeV}, below which they tend to interact via ionization rather than Bremsstrahlung. The resulting cascade of electrons, positrons, and gamma rays is called an extended air shower. Hadronic cosmic rays (protons and nuclei) can also initiate similar particle cascades when they collide with nuclei in the atmosphere, producing pions and other mesons, nucleons, and fragments of the original nuclei. Hadronic air showers may also contain muons, neutrinos, and electro-magnetic sub-showers from meson decays. Hadronic air showers make up the main background for ground-based gamma-ray telescopes. For more detailed information see e.g. \cite{grieder2010,Zyla:2020zbs}.

The mean interaction length in air is given by $\chi_0 = \SI{37}{g\per\centi\metre\squared}$ for Bremsstrahlung and $\frac{9}{7}\chi_0$ for pair production. For reference, the column density at sea level corresponds to about \SI{1000}{g\per\centi\metre\squared}. In a simple approximation (Heitler model), the number of particles (including gamma rays) in the air-shower doubles every interaction length, whereas the average energy per particle is halved. Thus, a \SI{100}{\GeV} gamma-ray photon could produce ten generations of particles whereas a \SI{100}{TeV} gamma-ray photon would produce twenty generations.

There are multiple methods for detecting air showers at the ground. The two most important ones for gamma-ray astronomy are the imaging air Cherenkov technique and particle detector arrays.

Imaging air Cherenkov telescopes utilize the fact that charged particles in air showers can emit Cherenkov light in the atmosphere if their speed $v$ exceeds the local speed of light.  The amount of energy emitted at frequency $\omega$ per unit length is given by $\frac{d^2E}{dx \, d\omega} = \frac{q^2}{4 \pi} \mu(\omega) \omega \left(1 - \frac{c^2} {v^2 n^2(\omega)}\right)$, where $q$ is the particle's charge and $n\left(\omega\right)$ and $\mu\left(\omega\right)$ are the (frequency-dependent) index of refraction and permeability of air. The refractivity $n\left(\omega\right)$ is approximately proportional to the density of the atmosphere, but also depends on the humidity. The Cherenkov spectrum in air peaks at blue or UV frequencies.

Cherenkov light is emitted at an angle of $\theta_C = \cos^{-1}\left(\frac{1}{n\left(\omega\right)}\right)$. The Cherenkov emission in air is strongly beamed forwards, with a maximum angle of \ang{2} near sea level. Cherenkov telescopes are set up to focus the Cherenkov light onto a very sensitive camera, typically consisting of hundreds or thousands of photo-multiplier tubes (PMTs) or silicon photo-multipliers (SiPMs). The shape and brightness of these images are used to reconstruct the direction and energy of the primary particle that initiated the shower.

Particle detector arrays detect the charged components of air showers directly. They typically consist of water tanks or scintillator plates, instrumented with PMTs which detect Cherenkov emission from charged particles crossing through the detector. The timing and the strength of the signal are used to reconstruct the direction and energy of the primary particle.


\subsubsection{Particle interactions in the detector volume}
\emph{Gamma rays} predominantly interact with matter in three different ways, with the dominant mechanism depending on the photon's energy and the material. Above tens of MeV, \emph{pair production} generally dominates: the photon disappears and an electron-positron pair is produced in the electric field of an atomic nucleus or electron. If the electron or positron have sufficient energy, they can emit further gamma rays which can in turn produce more electron-positron pairs. These cascades develop similar to the air showers described above, but on smaller length scales in solid media compared to the atmosphere. 

Between tens of keV and tens of MeV, the dominant process is \emph{Compton scattering}, in which the gamma ray transfers some of its energy to an electron and is deflected by an angle $\phi$. Assuming the energy loss of the photon  is much larger than the ionization energy, the energy of the scattered photon is given by $E_{\gamma^\prime} = \frac{E_\gamma}{1 + (E_\gamma/m_e c^2)(1-\cos\phi)}$, where $E_\gamma$ is the initial photon energy and $m_e$ is the electron mass.

For energies below tens of keV, the \emph{photoelectric effect} is the dominant interaction mechanism, in which the photon's energy is completely absorbed by an electron in an atomic shell, releasing the electron. Any energy in excess of the binding energy is converted to kinetic energy of the electron.

\emph{Charged particles} such as electrons, muons, and protons mainly interact via \emph{ionization}. Electrons (and positrons) with energies above some tens of MeV emit \emph{Bremsstrahlung} in the presence of nuclear electric fields, emitting gamma-ray photons and, if they are sufficiently energetic, initiating an electromagnetic cascade as described above. Protons, neutrons, and nuclei of sufficient energy may interact with other nuclei via the strong interaction, producing pions and other mesons in a \emph{hadronic cascade}, which proceeds similarly to the electromagnetic showers described above.

\subsection{Detector readout}
Simulating the detector readout adds realism to the simulated events. The resolutions of the measured quantities, such as energy, position, and timing, affect the event reconstruction and the overall instrument performance. The position resolution is generally the size of the instrument segmentation; e.g. the size of a single detector, strip, or pixel. The energy and timing resolutions are dependent on both the detector properties, such as material and size, as well as the readout electronics. The energy and timing measurements are also discretized when the signals are digitized (e.g. with an analog to digital converter).

Additional detector readout effects, including but not limited to dead time, thresholds, and trigger conditions, also affect the overall instrument performance. Dead time can affect overall count rates and, depending on the trigger system, lead to incompletely detected events. For example, consider the case in which the multiple detectors that make up an instrument trigger independently of each other. If an event occurs while the readout of a single detector is processing the previous event, it is possible that the event scatters in the dead detector and thus not all of its deposited energy is detected. Similarly, sub-threshold interactions, in which a particle deposits energy below the threshold of the readout element, can lead to undetected or incompletely detected events. Other common electronics effects, such as cross talk, can distort the measured parameters.

Incompletely detected events and distortions of the measured parameters due to realistic resolutions and electronics effects will affect the event reconstruction, and consequently, the final instrument performance. Additionally, an accurate simulated count rate is essential for estimating the flux of astrophysical sources. The detector readout is thus a key component of simulations. These effects are generally measured and modeled during the instrument calibration, and then incorporated into the simulation pipeline.

\subsection{Event reconstruction}

\begin{figure}
\centering
\includegraphics[width=0.49\linewidth]{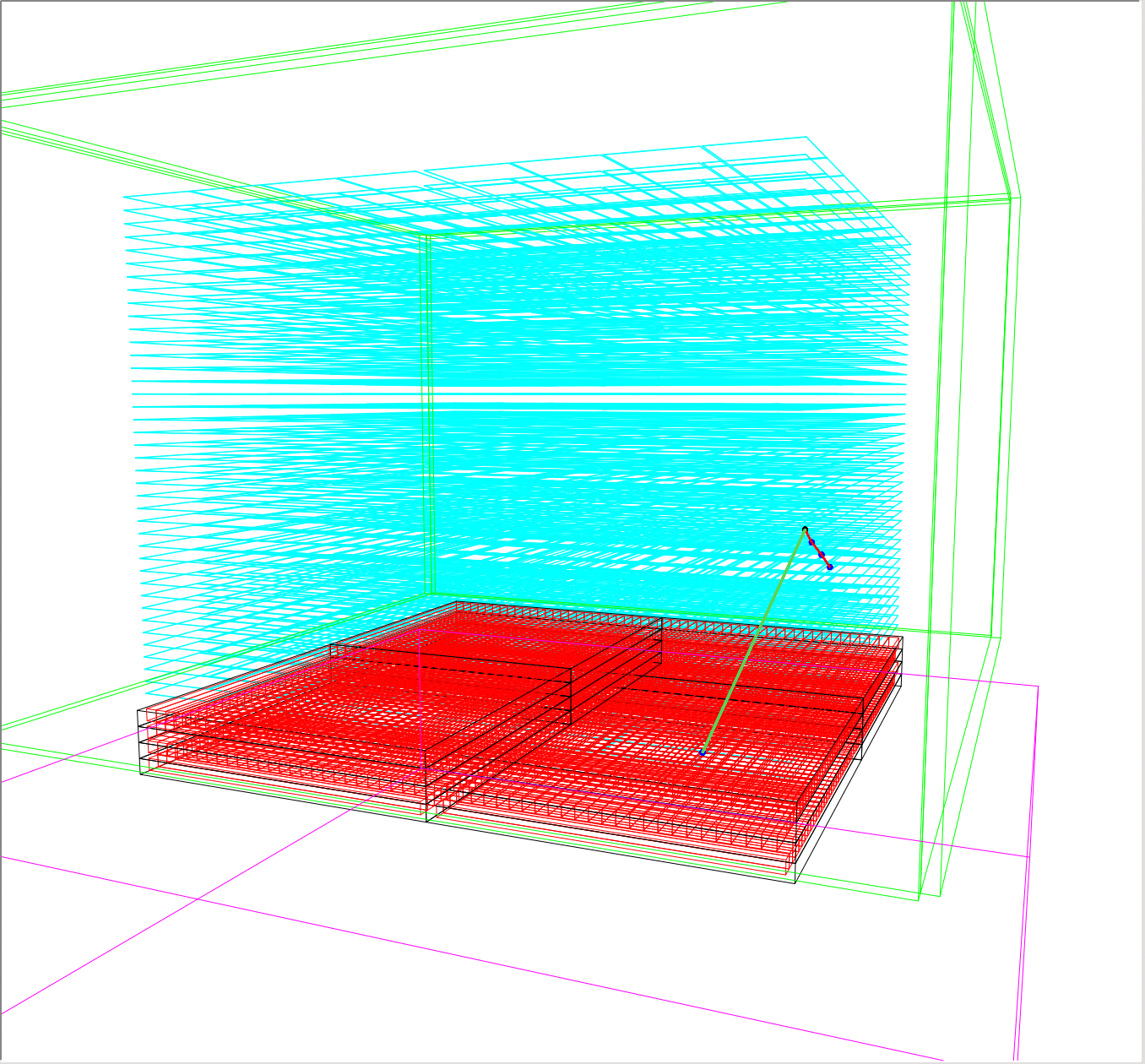}
\caption{A simulated gamma-ray photon interacting in the proposed {\it AMEGO-X}\cite{Fleischhack:2021mhc} detector via Compton interaction. The gamma ray crossed the top half of the tracker (in cyan) without interaction before Compton scattering. The scattered electron (blue track) was detected in four tracker layers before being absorbed. The scattered photon continued to travel through the tracker without interaction (green line) and deposited the remainder of its energy in the calorimeter (red) via a photoelectric effect interaction. Image produced with the MEGAlib simulation framework and previously published in \cite{Fleischhack:2021mhc}. \textcopyright{} H. Fleischhack, reproduced with permission.}
\label{fig:compton_event} 
\end{figure}

\begin{figure}
\centering
\includegraphics[width=0.99\linewidth]{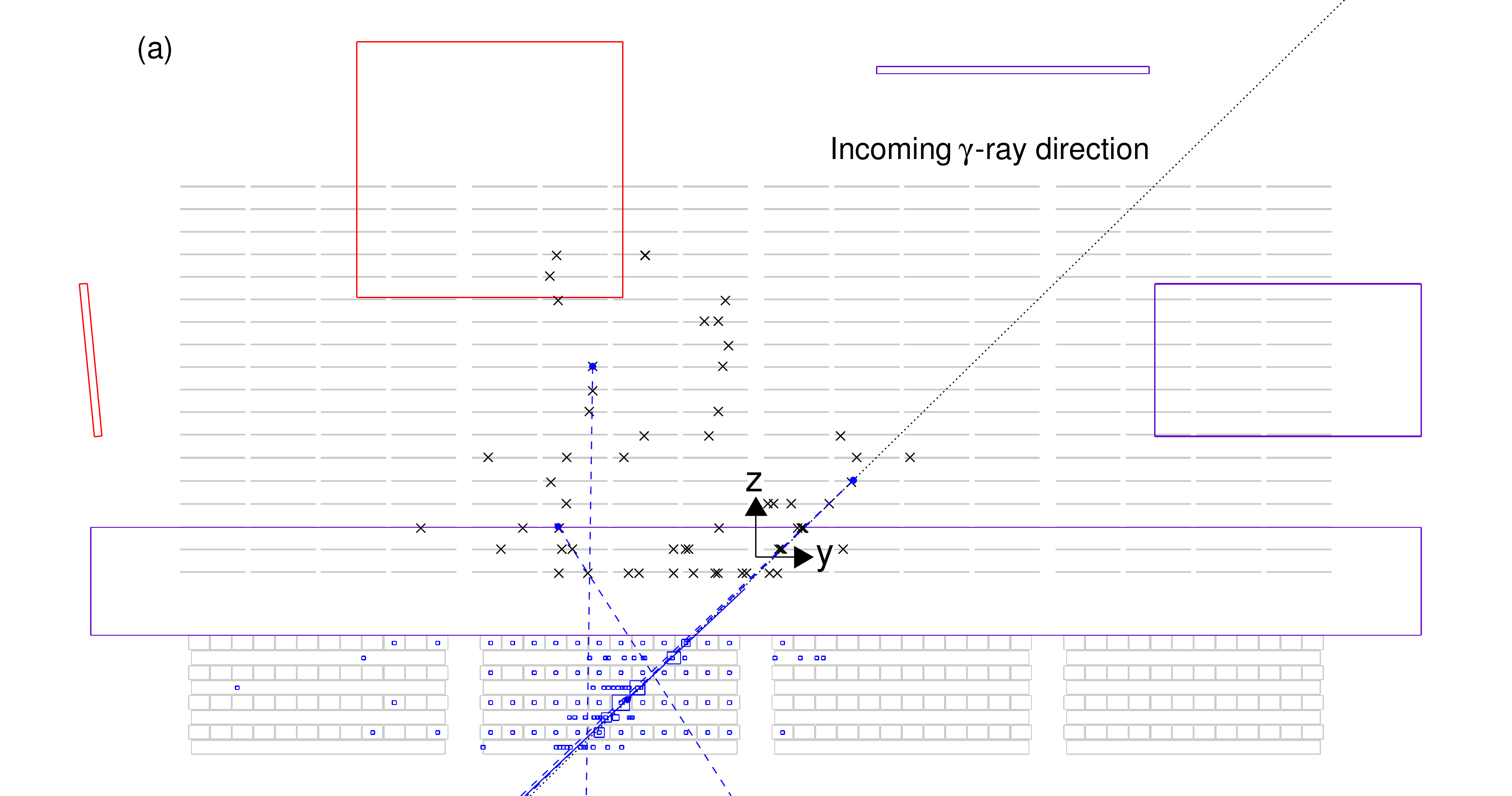}
\caption{A simulated {\it Fermi}-LAT Event.   The simulated gamma ray had 27 GeV of energy.
The small crosses represent the clusters in the silicon tracker,
while the variable-size squares indicate the reconstructed location and magnitude of the
energy deposition for every hit crystal in the calorimeter. The dotted line represents the true gamma-ray
direction.   The "backsplash'' from the calorimeter shower generates tens of hits in the tracker, 
and a few hits in the anti-coincidence detector  (colored boxes), which, however, are away from the direction
extrapolation and therefore do not compromise our ability to correctly classify the event as
a gamma ray.  This figure appeared as Fig. 12(a) of Ref.\cite{2012ApJS..203....4A}; reproduced by permission of the AAS.}
\label{fig:fermi_event} 
\end{figure}

\begin{figure}
\centering
\includegraphics[width=0.49\linewidth]{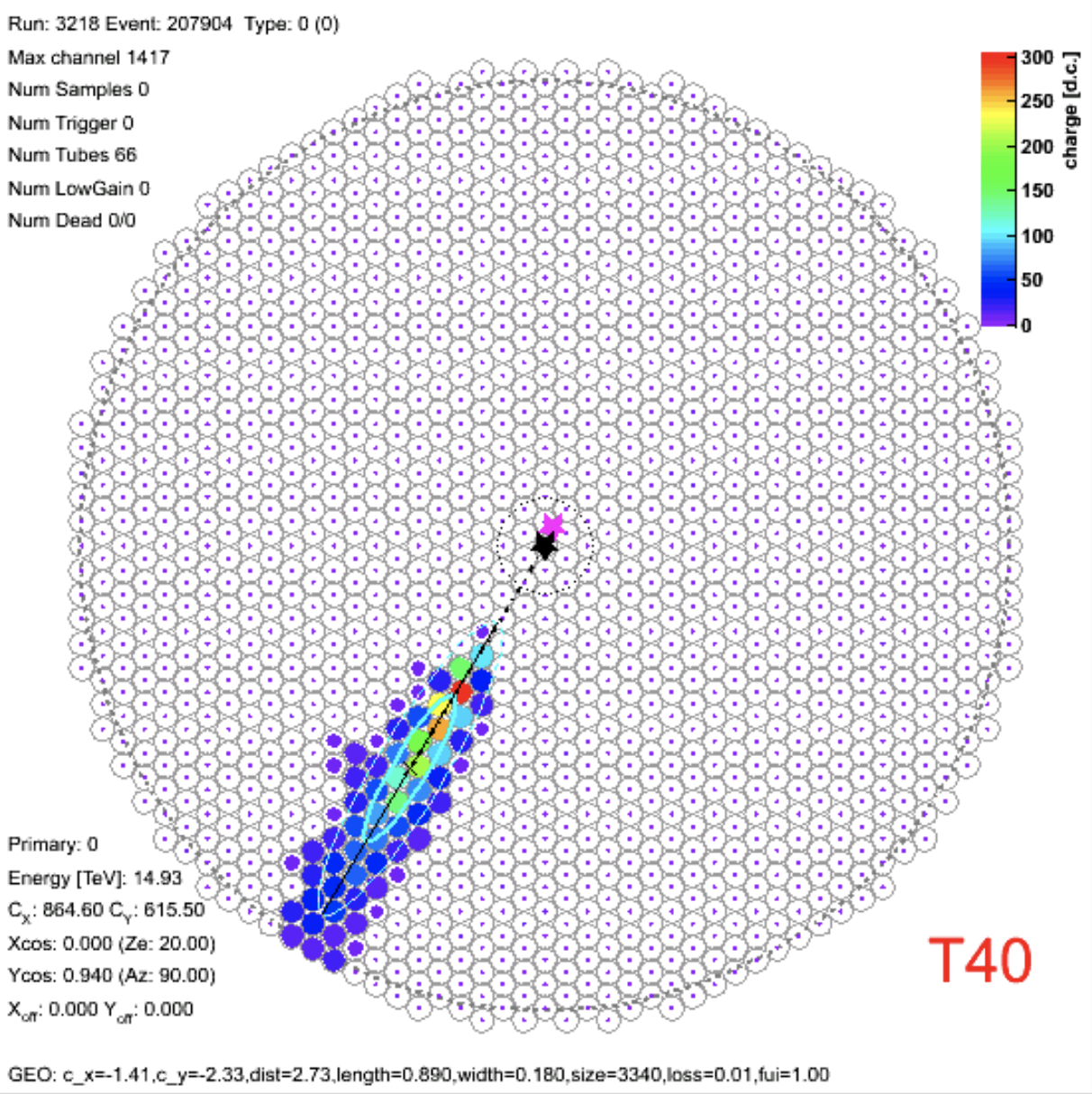}
\includegraphics[width=0.49\linewidth]{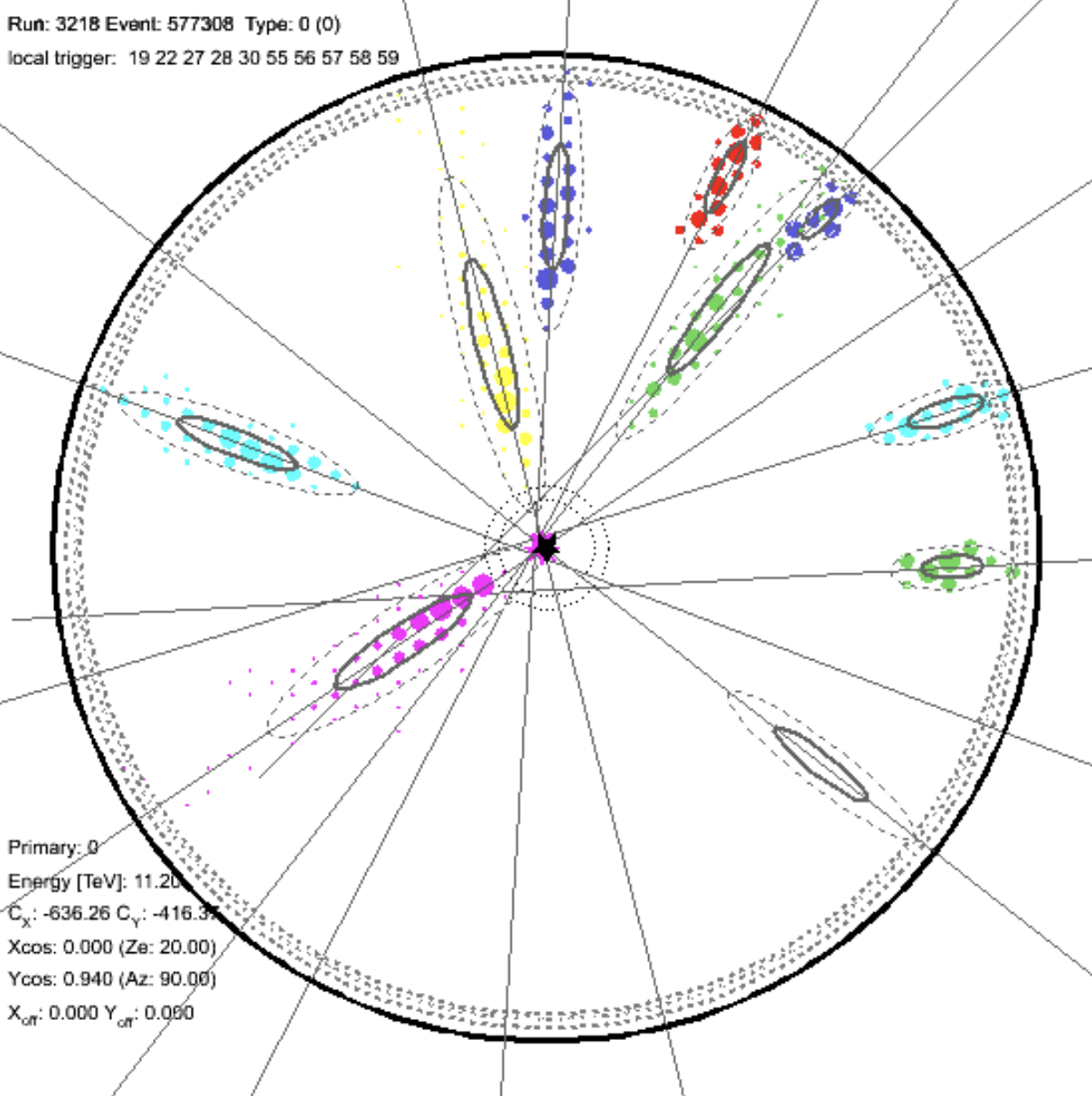}
\caption{Simulated image of a gamma-ray air shower in the CTA camera. The simulated gamma-ray source (black star) was located in the center of the camera. The reconstructed photon direction is marked with the purple star. Left: Single-telescope view. Colors correspond to the measured charge in each camera pixel (PMT/SiPM). Medium and large dots indicate that the pixel in question has passed ``image cleaning'' (removal of pixels containing just noise) and is being used for the reconstruction. Right: Multiple images of the same shower seen by different telescopes in the array, all pointing in the same direction. Colors correspond to the telescope ID, dot sizes to the measured charge. Images previously published in \cite{Maier:2017wzr}. \textcopyright{} G. Maier and J. Holder, reproduced with permission.}
\label{fig:cta_event} 
\end{figure}

\begin{figure}
\centering
\includegraphics[width=0.49\linewidth]{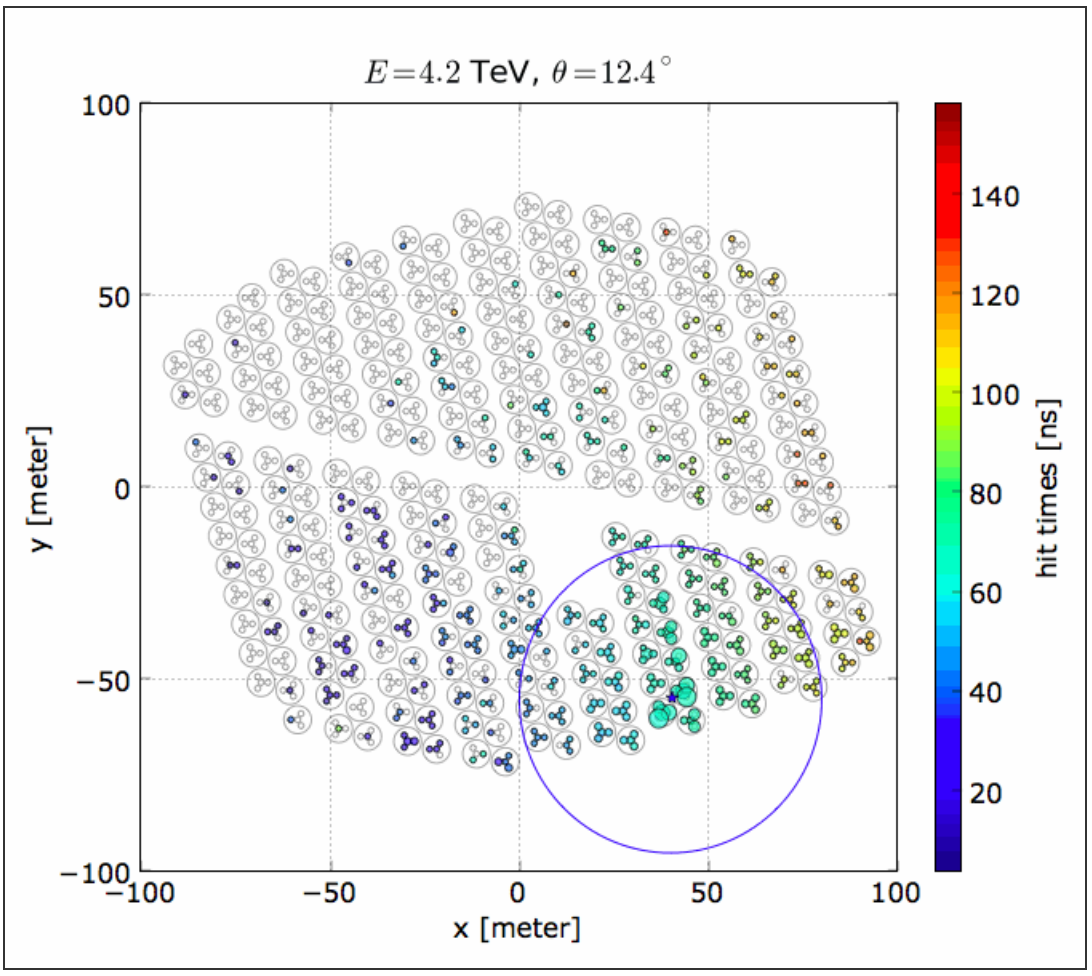}
\caption{A simulated gamma-ray shower seen by the HAWC detector. Each large circle corresponds to one water tank, with the smaller circles depicting the four PMTs in each tank. PMTs detecting light are marked by colored circles. The color corresponds to the hit time (used to reconstruct the shower direction) and the size of each circle corresponds to the amount of light seen (used to reconstruct the gamma-ray energy). Image from \url{https://www.hawc-observatory.org/observatory/ghsep.php}, reproduced with permission.}
\label{fig:hawc_event} 
\end{figure}

Gamma-ray telescopes typically record many signals for each incident particle.  This number can range from a few, in the case of relatively low energy Compton interactions, to thousands, it the case of very high energy events.   The task of the ``event reconstruction'' is to extract information about the incoming particle from all of those signals.  Figs.~\ref{fig:compton_event}, ~\ref{fig:fermi_event}, \ref{fig:cta_event} and \ref{fig:hawc_event} show examples of the single event information that is available for different types of gamma-ray telescopes (Compton, pair, IACT, and particle detector array, respectively).   As far as simulations are concerned, the key point is to have enough fidelity and granularity to develop and optimize the event reconstruction algorithms.

In the later three cases, the result of the event reconstruction includes estimates of the direction and energy of the incoming particle, as well as ``quality'' estimators that indicate if the event is more or less likely to be a gamma ray or cosmic-ray background.

In the case of Compton event reconstruction,  it is typically not possible to fully identify the direction of the incoming particle; rather the event reconstruction identifies a cone around the incoming direction.  The opening angle of the cone is given by the Compton scatter angle, and can be very large.   Further analysis of Compton events is often done in the ``Compton'' data space~\cite{Zoglauer2021}, which is defined by the astrophysical coordinates of the vector joining the first Compton scatter and the second interaction, and the Compton scatter angle as estimated from the kinematics of the event.  Additionally, Compton event reconstruction does typically estimate the energy and the event ``quality''.

\subsection{High-level data analysis}

High-level data analysis typically takes as input \emph{reduced data}, such as lists of photons or photon candidates with associated information from the event reconstruction such as arrival direction, arrival time, energy, and potentially other parameters measuring reconstruction quality or ``gamma-ness''. The input can also be binned data such as sky-maps, typically binned in Right Ascension (RA or $\alpha$) and Declination (DEC or $\delta$) or in Galactic latitude ($b$) and longitude ($l$) according to the photon's arrival direction, plus further bins according to energy, arrival time, etc. Instrument response files (effective areas, angular resolution, etc.), which are generated from simulations, are needed as well. The output of the high-level analysis generally consists of results such as the location(s), detection significance(s), morphology, and energy spectrum/spectra of detected source(s).

In general, most of the field uses either maximum likelihood estimation or Bayesian sampling of the posterior distribution. The likelihood is derived from comparing the detected number of photon counts per bin to a model of the source(s) convolved with the instrument response. Model parameters such as source location and the flux normalization of each source are varied to minimize the likelihood.

Many tools for high-level analyses of this kind exist, typically specific to an instrument or data format \cite[e.g.,][]{2019ascl.soft05011F, ZOGLAUER2006629, Maier:2017wzr}. Several efforts exist that allow simultaneous likelihood fits to data from multiple instruments, either by unifying the data format \cite{gammapy:2017} or by allowing encapsulating data and instrument response access into plugins, allowing the joint analysis of data in disparate formats \cite{Vianello:2015wwa}. Many of these packages include tools for high-level simulation, e.g. to generate synthetic measured spectra or skymaps given a ``true'' spectrum and instrument response.

\subsection{Performance metrics}
\label{subsec:performance_metrics}
Standard metrics of instrument performance are useful in comparing instruments to each other and completing trade studies when designing a new instrument. The main performance metrics are the instrument energy resolution, angular resolution, effective area, and sensitivity. In addition, the minimum detectable polarization is a useful metric for instruments with polarimetry capabilities. These metrics are generally dependent on the photon energy and the source's location in the instrument's field of view, and (for ground-based instruments) the elevation angle under which a source is observed. The performance is affected by any imperfections in the event calibration and event reconstruction pipelines.

The energy resolution describes how well an instrument can measure the energy of an incident photon, and can be determined by measuring the width of the Gaussian lines that occur when a (real or simulated) monoenergetic source is incident upon the instrument. The full instrument energy resolution differs from that of the individual elements, such as a single detector pixel or strip, if most gamma rays interact multiple times within the instrument. For ground-based instruments, the photon energy is reconstructed from either the Cherenkov light or the small part of the air shower that reaches the detector. Since the energies and detector sizes involved make beam tests impractical, both the energy reconstruction and the resolution measurements are entirely based on simulations. It is thus important to not just model the detector, but also the atmospheric profile and the shower development well to understand the energy resolution and other performance metrics.

The angular resolution describes the instrument's ability to spatially resolve sources on the sky. The angular resolution is determined by the point spread function (PSF), and is often defined as the angle in which 68\% of events are contained. For pair telescopes and ground-based instruments, the PSF is the probability distribution function of the offsets between the true $\hat{\nu}$ and reconstructed $\hat{\nu}'$ directions of each gamma ray. In the case of a Compton telescope, which localizes each photon to an event circle on the sky, the PSF is a distribution of the angular resolution measure (ARM): the ARM of each event is defined as the smallest angular distance from the event circle to the true source position. The shape of the PSF varies based on the incident gamma ray energy and the source location in the instrument's field of view. The angular resolution can be determined either from simulations or from measurements (e.g. beam test, radioactive lab sources, observations of known point-like sources).

The effective area is the area of a perfectly efficient ideal detector that detects the same number of counts as the real instrument, and is computed by multiplying the instrument efficiency by the cross-sectional geometric collecting area (for direct detection instruments) or the instrument area folded with the footprint of a gamma-ray air shower (for ground-based instruments). Thus, the effective area is a measure of instrument size and efficiency, which varies depending on the incident energy and source location in the instrument reference frame.

Since linearly polarized gamma rays exhibit sinusoidal modulation in the azimuthal Compton scatter angle, Compton telescopes can detect polarization by measuring this modulation (see e.g. \cite{Lei97} for an in-depth discussion of gamma ray polarimetry). The minimum detectable polarization (MDP) is the degree of linear polarization above which a statistically significant detection can be made, and is commonly used as a metric of the polarimetric performance. The MDP is a function of the source and background count rates, the observation time, and the modulation factor of a 100\% polarized beam (the modulation factor is the amplitude divided by the offset of the measured sinusoidal modulation) \cite{Weisskopf2010}. The modulation factor used in the MDP is often determined from simulations, since producing a fully polarized beam in the lab is non-trivial. The source and background rates are dependent on the source properties and the instrument effective area.

\subsection{Sensitivity estimates}

An extremely useful figure of merit is the ``flux sensitivity'', i.e., the minimum flux that is likely to result in a statistically significant source detection given a fixed observing program. In short, this requires detecting a source above a background.  

The flux sensitivity can always be estimated with an exhaustive simulation of a ``grid'' of sources with varying fluxes and spectral indices generated on top of simulated backgrounds, which are then passed through the standard event reconstruction and high-level analysis chain to determine which fluxes and indices are detectable.    The approach is also used with ``fake source injection'' techniques, wherein sources with known properties are injected into copies of the real data; this latter approach better captures both the sky backgrounds and instrumental artifacts.

However, it is often possible to get a quite accurate estimate of the flux sensitivity with far less work.   Several methods exist to do this, and they are typically enhancements of ``aperture photometry'', i.e., counting the photons that lie within an aperture defined by the PSF containment radius, and using that to estimate the flux needed to establish a significant excess above the expected background level.  Typically, the enhancements involve weighting the backgrounds to discount events that are further away from the source position, or have energies that are less likely to be from the source.  In practice this involves 1.) estimating  ``effective background'' at the source location by convolving the flux model of the background with the IRFs,  2.) determining the number of counts from the source ($n_{\rm src, detect}$) that would be needed to establish a detection at a given statistical significance threshold, 3.) converting $n_{\rm src, detect}$ back to the minimum detectable flux given the instrument response functions (IRFs, see \S\ref{subsec:IRFs}) and the observing program in question.

An example of such a method is described in Appendix A of Ref.~\cite{2FGL}, where it is also used to provide an estimate of the source localization power.


\section{Simulation tools for different types of instruments}

\subsection{Simulating energy deposition in the instrument}

\subsubsection{Air shower simulations}

There are several software packages that simulate extensive air showers, the most commonly used one being the CORSIKA framework \cite{1998cmcc.book.....H}. Air shower physics are complex and the simulations require inputs from quantum electrondynamics (QED), quantum chromodynamics (QCD), nuclear physics, classical electrodynamics, and optics. Additionally, good models of the atmospheric properties (gas content, pressure, optical properties) are needed. Simulation steps include the simulation of hard (high-energy) interactions between the primary particle (or energetic secondary particles) and the atmosphere, production (and decay, if applicable) of secondary particles, the energy loss of charged particles via ionization and other mechanisms, deflection of charged particles in the Earth's magnetic field, and (if required) the emission of Cherenkov light, fluorescence light, or radio waves.

Pure QED and QCD calculations are prohibitive due to convergence issues. Instead, various (interfaces to) external models have been implemented to treat high-energy hadronic \cite[e.g.,][]{OSTAPCHENKO2006143, Riehn:2017mfm, Werner:2007vd}, low-energy hadronic \cite[e.g.,][]{Fesefeldt:162911,BATTISTONI201510,1998}, and electromagnetic interactions \cite[e.g.,][]{egsnrc}. These codes are typically tuned to match measurements at particle colliders and from air shower experiments. The users can pick which interaction model to use. The differences between predictions from different interaction models are a measure for the systematic uncertainty of these codes. The uncertainties can be pretty significant, especially for hadronic interactions and high energies. For ground-based gamma ray observatories, it is often preferable to measure the hadronic background instead of relying completely on simulations.

For particle detector arrays such as HAWC and Tibet-ASgamma, the optical properties of the atmosphere are unimportant as the data analysis only relies on a prediction of how many charged particles will deposit energy in the detectors. The shower development does depend on the density profile, but to first order it is enough to know the total mass overburden, which can be derived from a simple measurement of the air pressure at the detector site. Conversely, IACTs rely on the amount of Cherenkov light reaching the detector, which requires detailed knowledge of the density profile and composition of the atmosphere. Both the intensity and the angle of the emitted Cherenkov light depend on the index of refraction at the position where the light is emitted, and the light loss due to scattering and absorption depends on the composition of the atmosphere (including its aerosol content) down to the telescope site. See \cite{BERNLOHR2000255} for details. CORSIKA includes several atmospheric models appropriate for various latitudes. The users can also supply their own.

In many cases, it is appropriate to approximate the Earth's surface as a plane, rather than taking its curvature into account. This speeds up air shower simulations significantly. However, for the simulation of very inclined showers it is necessary to model the atmosphere as a spherical shell.

Another way to reduce computing time and memory usage is to follow only a representative set of shower particles, where each representative is a stand-in for many real particles. This method is called ``thinning''. Thinning may introduce artificial fluctuations or graininess into the simulated shower profile. The equivalent for Cherenkov photons is also referred to as  ``bunching''. The bunch size is typically chosen such that each bunch produces on average one photo-electron in the camera, after losses due to atmospheric absorption, inefficient mirrors etc. are taken into account.


\subsubsection{Ray tracing}

Dedicated ray-tracing simulations are performed to follow the path of Cherenkov photons through the optical structure of Cherenkov telescopes. Optical elements of Cherenkov telescopes typically comprise segmented spherical or parabolic mirrors, as well as baffles or Winston cones around the camera pixels to limit background light. Ray-tracing simulations should take into account the (wavelength-dependent) reflectivity of the mirrors and Winston cones (if installed), the optical point spread function and alignment of each mirror facet, and losses due to gaps between mirrors or shadowing by the camera or support structure. Ray-tracing can be performed by dedicated software such as ROBAST \cite{robast} used e.g. by Cherenkov Telescope Array (CTA), or as part of telescope-specific simulations packages such as  sim\_telarray \cite{Bernloehr2008} and matelsim \cite{matelsim}. Ray tracing can either be performed for each simulated gamma ray event separately, or ray-tracing simulations can be used to derive effective optical point spread functions and efficiencies which can then be folded into the simulated gamma ray images on the camera.

\subsubsection{Simulating particles in matter with Geant4}

The Geant4~\cite{AGOSTINELLI2003250,ALLISON2016186,1610988} simulation toolkit is the standard tool used in high-energy particle physics to simulate the passage of particles through matter, as well as the energy that the particles leave in active detectors.  In summary, Geant4 propagates particles through a detailed spatial model of a detector, keeping track of particle interactions and decays, and propagating any resulting particles until they exit the detector volume, decay or are otherwise destroyed in an interaction, or come to rest because of interaction induced energy losses.  Geant4 then produces a list of the resulting particles and their interactions with the detector model (i.e., the identity of the interactive particle, the position of the interaction, and the amount of energy deposited).  At that point it is up to the users to model the detector response to the deposited energy, i.e., to simulate the resulting digitized signals that would result from the deposited energy.

Geant4 is used extensively as a front-end by the MEGALib package (see \S\ref{subsec:det_elec}), by SoftWare for Optimization of Radiation Detectors (SWORD) \cite{Sword}, and by GlastRelease, the software framework used to simulate and reconstruct the interactions of individual gamma rays with the {\it Fermi}-LAT (again, see \S\ref{subsec:det_elec}).   In practice, Geant4 is the standard software package for simulating the interactions of high-energy particles with meter-scale instruments (as opposed simulations of much larger air-showers, which are done with the other tools described in this section).

\subsection{Simulating detector electronics}\label{subsec:det_elec}

The tools used to simulate detector electronics vary widely across experiments. Often the detector electronics effects are incorporated into the simulations with instrument-specific tools. In the case of {\it Fermi}-LAT, for example, the same packages that calibrate the events are used to simulate the detector readout; these packages are part of the {\it Fermi} software tools\footnote{https://github.com/fermi-lat}. Detector electronics can also be incorporated into simulations via more general software tools: in particular, MEGAlib \cite{ZOGLAUER2006629} and sim\_telarray \cite{Bernloehr2008}.

GlastRelease, the suite of software packages use to simulate, reconstruct and analyze {\it Fermi}-LAT data, includes dedicated packages to simulate the detector electronics response of each of the instrument sub-systems.   These packages can use the instrument calibrations as input, in effect inverting the calibrations so as to map energy deposition to a digital signal level, and applying thresholding to determine which hits would actually be read out by the instrument.

MEGAlib is a software package designed for the simulation and analysis of Compton telescopes. MEGAlib includes energy resolution, depth resolution, trigger thresholds, and other detector electronics effects in the simulations. These parameters are specified in the mass model and can thus be easily incorporated in any post-simulation analysis. However, MEGAlib in its current form does not allow the user to vary electronics effects across a single detector. In other words, a standalone tool is needed to vary resolutions and trigger thresholds across the pixels or strips of a single detector, if such a variation is required. Since MEGAlib is open source, it is straightforward for such a standalone tool to interface with the MEGAlib simulations and mass model.

sim\_telarray is an extension to CORSIKA that simulates the detector electronics for IACTs. In addition to electronics effects, sim\_telarray can also simulate other instrumental and atmospheric effects.

\subsection{Event reconstruction}

The reconstruction of simulated events typically uses the same software and processing chain as measured data. In some cases, simulations require certain pre-processing steps. For example, the addition of detector electronics noise may be performed at the reconstruction stage, enabling the same simulation file to be used under different noise conditions.

\section{Trade studies and instrument design}

\subsection{Figures of merit and sensitivity metrics}

When designing an instrument, simulations are used to determine the projected performance. Figures of merit such as energy resolution, angular resolution, and effective area are useful in assessing the scientific potential of an instrument design. For example, improved angular resolution leads to improved source localization accuracy, and improved energy resolution can lead to gains in gamma ray line science capabilities.

The projected instrument sensitivity describes the minimum signal that an instrument can detect above the background, and is commonly compared to the sensitivities of already existing instruments. These sensitivity comparisons can be complex, since one must assume an observation time, a background rate, the energy spectrum of the source, and the effective area of the instrument. The background rate and effective area are often determined from simulations, but the effective area can change based on the source position in the instrument's field of view. Typical observation times are 1~Ms, 1~year, or the lifetime of the mission. Variations in the assumed observation times of the sensitivity curves from different instruments can make sensitivity comparisons non-trivial. However, it is not necessarily the best practice to assume the same observation time across instruments, since observing strategies and instrument fields of view differ. In other words, a pointed instrument with a narrow field of view will observe a source for less time than a wide field instrument, and an instrument cannot observe for longer than its lifetime.

\begin{figure}
\centering
\includegraphics[width=0.8\textwidth]{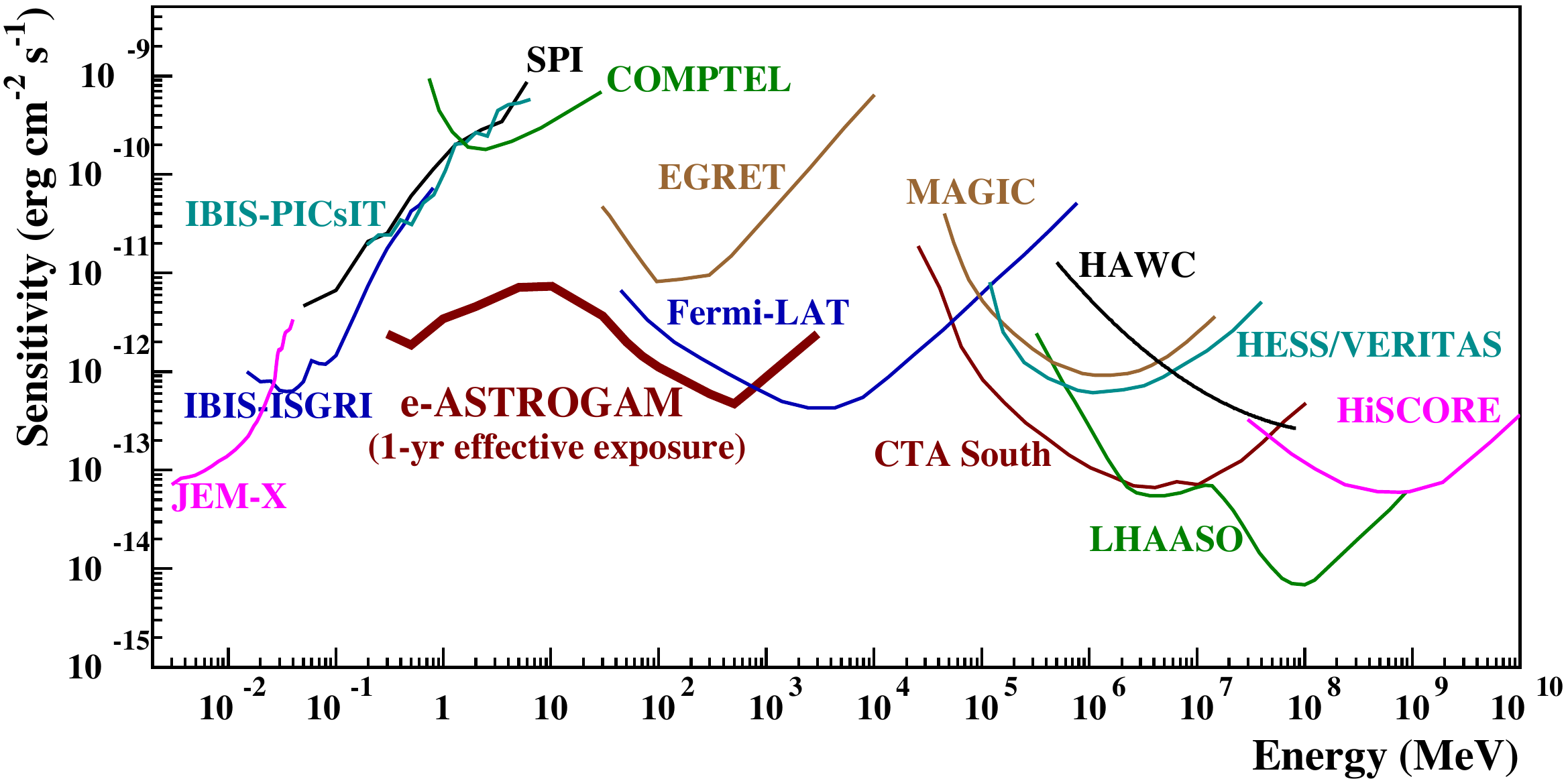}
\caption{A comparison of the continuum sensitivities of numerous gamma ray instruments, adapted from Figure 1.0.1 in Ref.~\cite{eAstrogam2}.
\label{fig:continuumsensitivity}}
\end{figure}

An example sensitivity comparison performed by the e-ASTROGAM collaboration is shown in Figure~\ref{fig:continuumsensitivity} \cite{eAstrogam2}. The e-ASTROGAM team assumed an observation time of 1~year to compute the 3$\sigma$ sensitivity.

\subsection{Examples of trade studies}

Trade studies are a crucial part of the instrument design process. The instrument design is optimized by varying several parameters and comparing the predicted performance as well as costs and potential risk factors of the different implementations. Simulations play an important role here as they are typically many orders of magnitude cheaper and faster than building detector prototypes.

In addition to science performance, trade studies also need to take into account various external factors such as differences in cost and timelines for development, procurement, and production. For ground-based instruments, accessibility of the site and other geographical constraints must be considered, whereas for space-based and balloon-borne instruments, limits on size, weight, and power are an important consideration.

Many trade studies are only conducted internally and not published in full. Some examples of trade studies are the comparison of the CTA site candidates \cite{HASSAN201776} and array layouts \cite{2017ICRC...35..811C}.

\section{Using simulations for science}

\subsection{IRFs: instrument response characterization}\label{subsec:IRFs}
Instrument response functions (IRFs) are parametrized representations of the instrument performance, and are needed to determine source properties from observations. For example, to perform spectral analysis of an astrophysical source, IRFs are used to extrapolate the emitted spectrum and flux of the source from the measured spectrum and count rate. IRFs are generally calculated from Monte Carlo simulations of gamma rays interacting with the instrument and often can be separated into three parts: the energy dispersion, the effective area, and the point spread function. Each of these three IRFs is usually dependent on the incident energy of the gamma ray, the incident direction of the gamma ray in the instrument reference frame, and the event selections used. The IRFs are typically parametrized in ``local" or ``instrument" coordinates. For a space-based instrument such as {\it Fermi}-LAT, local coordinates refer to the polar and azimuthal angle with respect to the instrument boresight. For ground-based instruments, local coordinates refer to the altitude (i.e. the amount of atmosphere above the instrument) and the azimuth.

\begin{figure}
\centering
\subfloat[]{\includegraphics[width=0.6\textwidth]{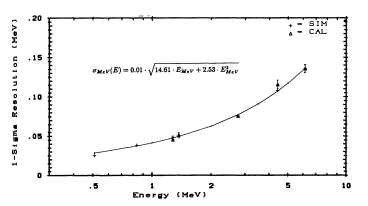}}
\\
\subfloat[]{\includegraphics[width=0.6\textwidth]{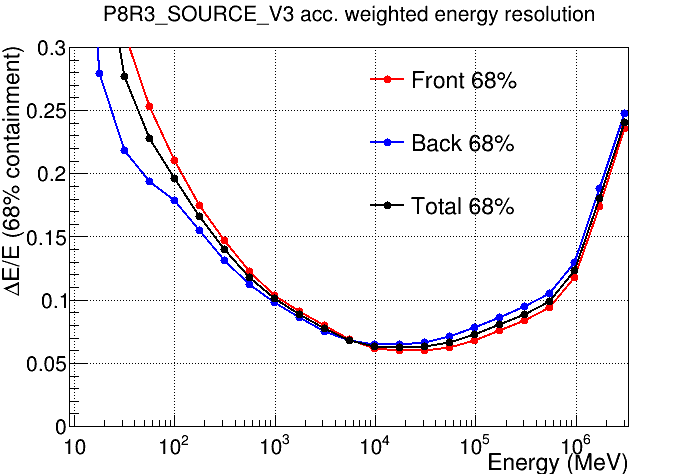}}
\\
\subfloat[]{\includegraphics[width=0.6\textwidth]{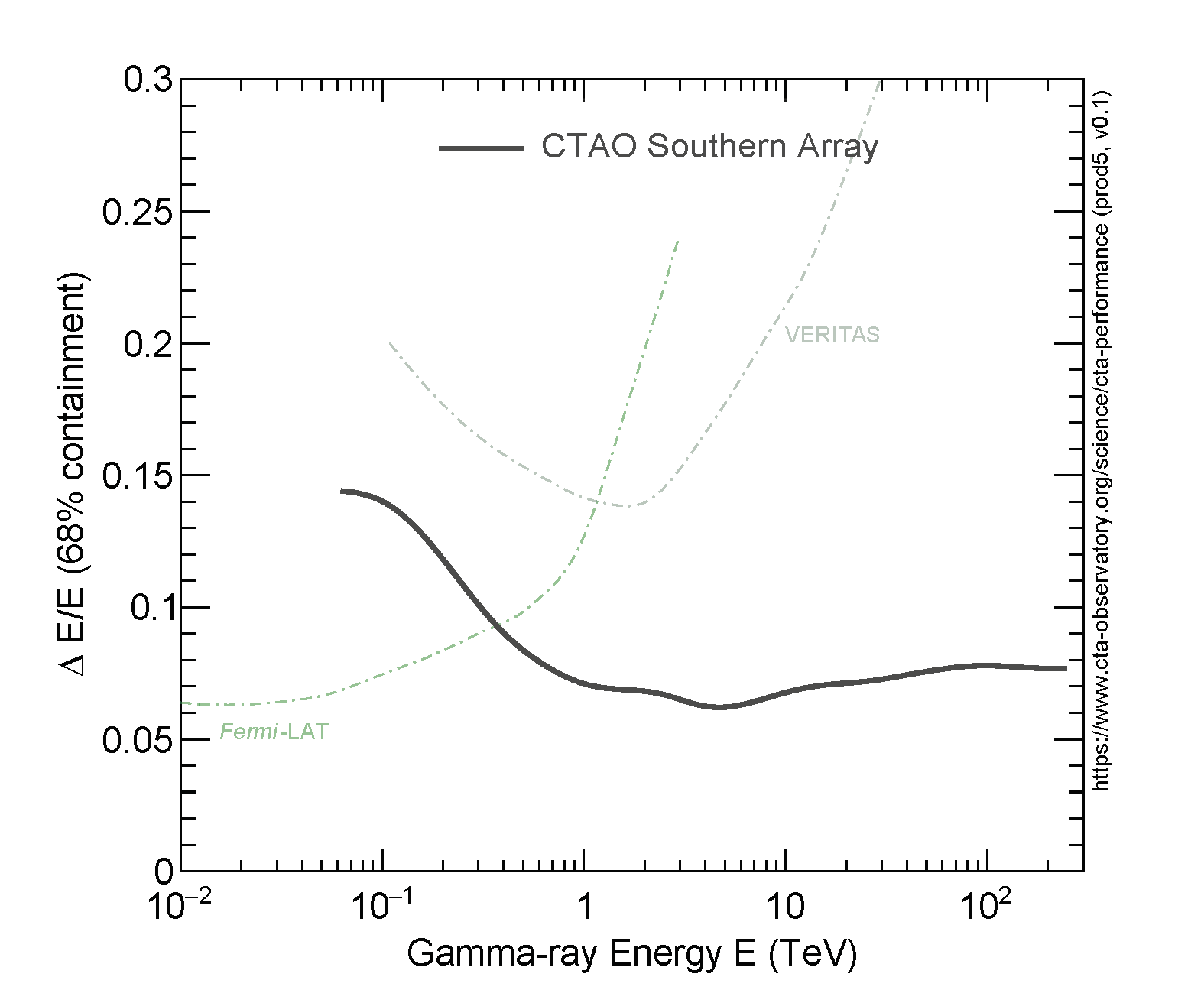}}
\caption{The energy resolution as a function of energy for (a) {\it CGRO}-COMPTEL \cite{Schonfelder93}, (b) {\it Fermi}-LAT \cite{LATperformance}, and (c) CTA \cite{CTAperformance}. The energy resolution is expressed as the width of the interval around the true energy which contains 68\% of the distribution. If the spectral lines are fit with a Gaussian distribution, the 68\% containment is equivalent to one Gaussian sigma, as in (a). In (b), ``front'' and ``back'' refer to the location of the pair conversion in the tracker. In (c), the energy resolution of CTA is compared to that of {\it Fermi}-LAT and VERITAS.  Fig.~(a) appeared as Fig.~30 of Ref.\cite{Schonfelder93}; reproduced by permission of the AAS. Fig.~(b) reproduced by permission of the {\it Fermi}-LAT collaboration. Fig.~(c) made use of the CTA instrument response functions provided by the CTA Consortium and Observatory, see https://www.cta-observatory.org/science/cta-performance/ (version prod5 v0.1; \cite{CTAperformance}) for more details.} 
    \label{fig:energydisp}
\end{figure}

The energy dispersion function, sometimes known as the energy redistribution function, describes the probability of measuring an event at energy $E'$ for a gamma ray with incident energy $E$. The instrument energy resolution is baked into the energy dispersion function as a spreading in measured energy $E'$ around true energy $E$. The energy dispersion function also incorporates other instrumental effects that skew the measured energy, such as low energy tailing due to incompletely absorbed events. Fig.~\ref{fig:energydisp} shows the energy resolution as a function of energy for Compton telescope aboard the Compton Gamma-Ray Observatory ({\it CRGO}-COMPTEL) (MeV energy range), {\it Fermi}-LAT (GeV energy range), and CTA (TeV energy range). The energy resolution is often conveniently used to summarize the information contained in the energy dispersion function, and is usually shown as a function of energy (as in Fig.~\ref{fig:energydisp}) or as a function of instrument boresight angle.

\begin{figure}
\centering
\subfloat[]{\includegraphics[width=0.6\textwidth]{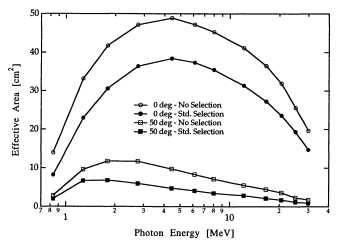}}
\\
\subfloat[]{\includegraphics[width=0.6\textwidth]{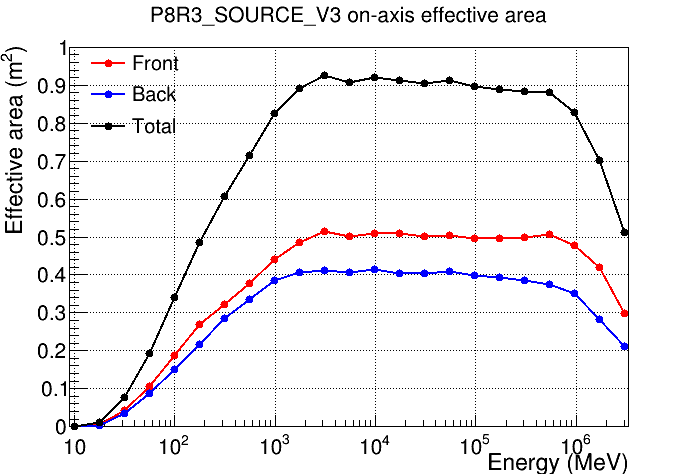}}
\\
\subfloat[]{\includegraphics[width=0.6\textwidth]{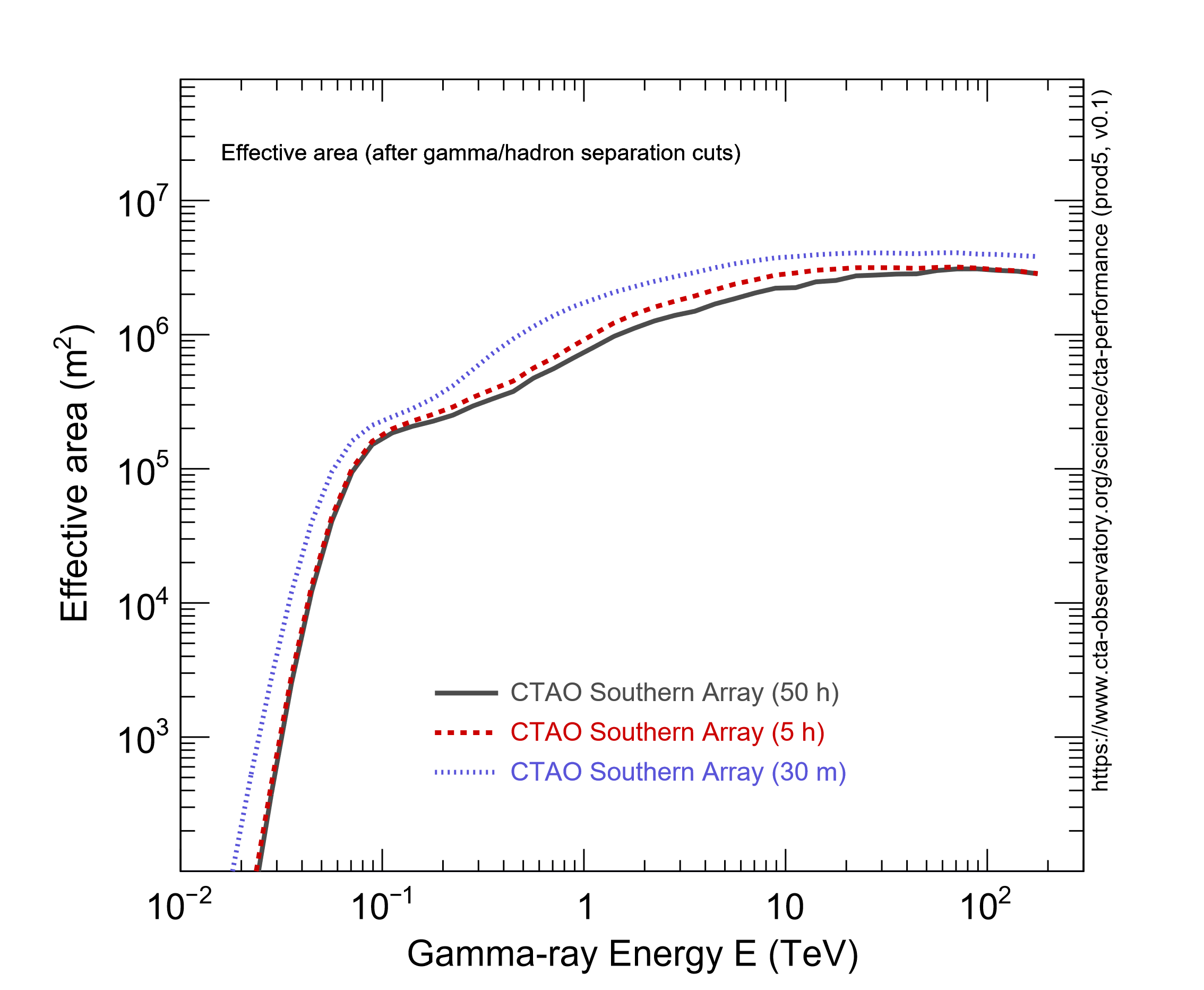}}
\caption{The effective area of (a) {\it CGRO}-COMPTEL \cite{Stacy96}, (b) {\it Fermi}-LAT \cite{LATperformance}, and (c) CTA \cite{CTAperformance} as a function of energy. (a) The effective area on-axis (0$^\circ$) and at 50$^\circ$ off-axis is shown, each with no event selections and the COMPTEL standard event selections. (b) The on-axis effective area, in which ``front" and ``back" refer to the conversion location of the pair event in the tracker. (c) The effective area for a variety of observation times, without an event cut in the reconstructed event direction.  Fig.~(a) appeared as Fig.~2 in \cite{Stacy96}, reproduced with permission \copyright ESO. Fig.~(b) reproduced by permission of the {\it Fermi}-LAT collaboration. Fig.~(c) made use of the CTA instrument response functions provided by the CTA Consortium and Observatory, see https://www.cta-observatory.org/science/cta-performance/ (version prod5 v0.1; \cite{CTAperformance}) for more details. \label{fig:effarea}}
\end{figure}

The effective area, introduced in \S\ref{subsec:performance_metrics}, is the area of an ideal instrument that detects the same number of photons as the real instrument. The effective area is required for measuring the flux of astrophysical sources. Fig.~\ref{fig:effarea} shows the on-axis effective area as a function of  energy for {\it CGRO}-COMPTEL, {\it Fermi}-LAT, and CTA. The effective area is also often plotted as a function of instrument boresight or azimuth at a single energy.

The point spread function (PSF) describes the instrument's imaging response. For pair telescopes and ground-based detectors (IACTs, water Cherenkov), the PSF is the probability density to reconstruct an incident direction $\hat{\nu}'$ for a gamma ray with incident direction $\hat{\nu}$. The PSF is usually taken to be radially symmetric with a central peak and more or less significant tails, and modeled for example as a Gaussian distribution or a King function \cite{1962AJ.....67..471K} (re-parametrized Student-t function \cite{10.2307/2331554}).

In the case of a Compton telescope, the PSF is more complex, since each photon is reconstructed to an event circle on the sky rather than a point. The PSF is defined in the Compton data space, a data space consisting of the initial Compton scatter angle $\phi$ and the direction of the scattered gamma ray in celestial coordinates $(\psi,\xi)$. In this data space, a point source creates a cone with an opening angle of $90^\circ$ and an apex at the point source position in celestial coordinates $(\psi_0,\xi_0)$; this cone is the PSF of a Compton telescope \cite{Zoglauer2021}. The thickness of the Compton data space cone describes the angular resolution, and the ARM described in \S\ref{subsec:performance_metrics} is a one-dimensional projection of this cone. The Compton imaging response describes the probability that an event emitted in a certain image space bin (with a particular energy) is detected in the Compton data space bin, and thus has at minimum 5 dimensions: the three data space dimensions representing the initial Compton scatter angle and the direction of the scattered gamma ray; and the two dimensional coordinates of the image space bin \cite{Zoglauer2021}. Two additional dimensions representing the emitted energy and the measured energy would ideally also be incorporated.

\begin{figure}
\centering
\subfloat[]{\includegraphics[width=0.6\textwidth]{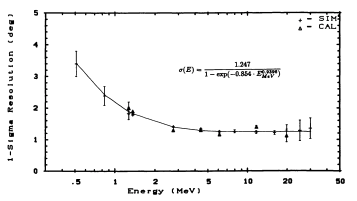}}
\\
\subfloat[]{\includegraphics[width=0.6\textwidth]{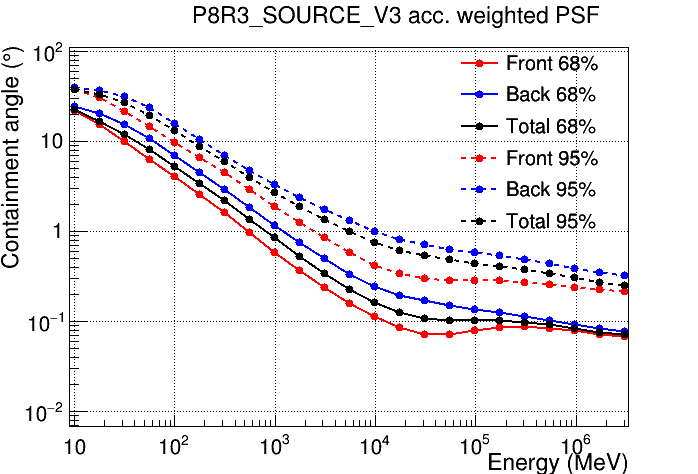}}
\\
\subfloat[]{\includegraphics[width=0.6\textwidth]{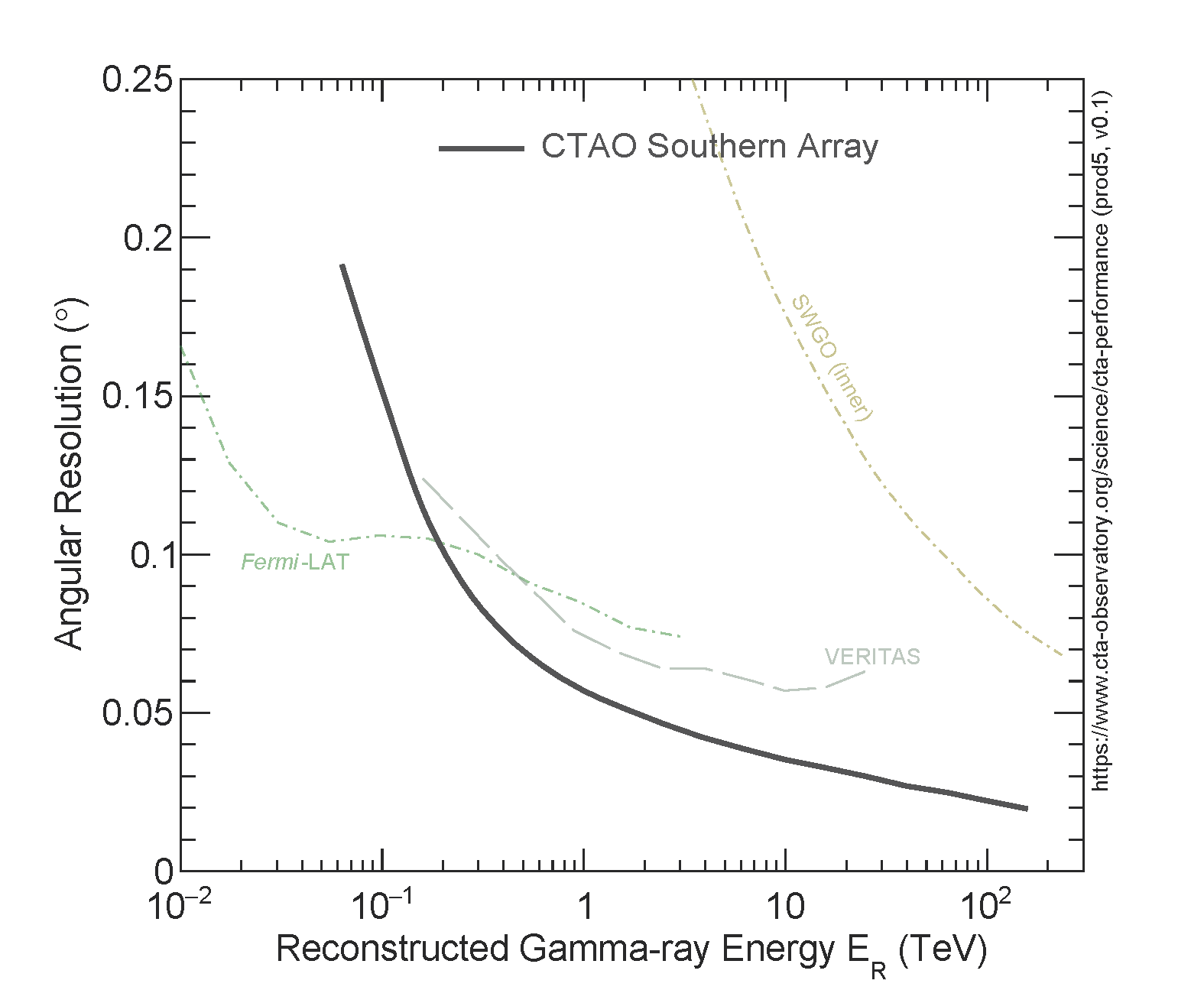}}
\caption{The angular resolution as a function of energy for (a) {\it CGRO}-COMPTEL \cite{Schonfelder93}, (b) {\it Fermi}-LAT \cite{LATperformance}, and (c) CTA \cite{CTAperformance}. (a) The one sigma width of the {\it CGRO}-COMPTEL ARM distribution. (b) The 68\% and 95\% containment angles of the {\it Fermi}-LAT PSF, in which ``front" and ``back" refer to the conversion location of the pair event in the tracker. (c) The 68\% containment angle of the CTA PSF, compared to a few other gamma ray instruments.     Fig.~(a) appeared as Fig.~28 in Ref.\cite{Schonfelder93}; reproduced by permission of the AAS. Fig.~(b) reproduced by permission of the {\it Fermi}-LAT collaboration. Fig.~(c) made use of the CTA instrument response functions provided by the CTA Consortium and Observatory, see https://www.cta-observatory.org/science/cta-performance/ (version prod5 v0.1; \cite{CTAperformance}) for more details.
\label{fig:psf}}
\end{figure}

The angular resolution as a function of energy is shown in Fig.~\ref{fig:psf} for {\it CGRO}-COMPTEL, {\it Fermi}-LAT, and CTA. Similarly to the energy resolution, the angular resolution can be used to summarize the information contained in the PSF, and is often defined as the angle in which 68\% of events are contained in the PSF. The angular resolution is often shown as a function of energy, instrument boresight, or azimuth.

\subsection{IRFs for variable observing conditions}

In addition to the energy and direction of the incoming photon, the instrument response of gamma ray detectors can depend significantly on various observing conditions as well as the state of the detector (e.g., aging of detector components, noisy or disabled sub-detectors).

IACTs are particularly sensitive to atmospheric conditions (seasonal variations in the density profiles, the presence of aerosols or clouds) and night sky background (e.g. stars or planets in the field of view and scattered moonlight, both of which can be either attenuated or further scattered towards the telescope by clouds). Mirror alignment and reflectivity as well as the gain of the photomultipliers degrades over time, but can be improved by periodically cleaning, re-coating, and re-aligning the mirrors and Winston cones and adjusting the PMT voltage. PMTs or even entire telescopes can be temporarily disabled due to stars in the field of view or hardware issues. Additionally, the three existing large IACT facilities H.E.S.S., MAGIC, and VERITAS have all undergone significant changes and hardware upgrades over the years \cite{Giavitto:2017ghe, Rajotte:2014pra, 2009arXiv0907.4826N,  2016_magic1, 2016_magic2}. Observations might also be performed under different trigger/voltage settings or with additional filters in front of the cameras.

There are two strategies of coping with these different observing conditions and hardware configurations. The first one involves simulating a large, multi-dimensional grid of atmospheric, observing, and instrument conditions, to cover all combinations occurring in data (by potentially interpolating between neighboring grid points). In this case, simulations can be performed independently of data taking, and data are ready to be analyzed as soon as they have been recorded. The downside is that the number of grid points and hence the needed resources (computing time, disk space) grow quickly as the number of dimensions increases, while only a small fraction of the simulated combinations may actually occur in reality. Another strategy employed e.g. by the H.E.S.S. collaboration is to produce dedicated simulations for each night's observing runs, following the actual position of the observed sources in the sky and simulating all other conditions as close to reality as possible \cite{Holler:2020duc}. The downside of this approach is that these dedicated simulations (potentially including cut optimization etc) have to be run after data-taking has concluded, before the data can be analyzed.

Particle detectors like HAWC are much less sensitive to changes in atmosphere and do not rely on the optical properties at all, thus not necessitating many different sets of instrument responses.

Balloon-borne gamma ray detectors, especially on long and ultra-long duration balloon flights, can be subject to atmospheric changes as they float around the Earth. In particular, changes in altitude affect the geomagnetic cutoff rigidity and thus the background count rate. Similarly to the IACT approach, one can include an altitude dimension in the IRFs -- in other words, compute a separate IRF for each altitude bin -- , as was done for the analysis of the COSI 2016 balloon flight data (e.g. \cite{Siegert2020}).

Satellite-based experiments like {\it Fermi}-LAT are less prone to degradation (not being exposed to the atmosphere) and once launched, the hardware is typically not upgraded. However, the rate of background hits caused by cosmic-ray events varies over the orbit (e.g. depending on proximity to the South Atlantic Anomaly region), which may necessitate separate response functions binned in noise rates.

\subsection{Fast simulations to characterize signal significance}

In general, the following are needed to be able characterize the statistical significance of a signal:
\begin{enumerate}
\item instrument response functions for gamma rays,
\item flux model for sky,
\item simulation of non-gamma ray backgrounds.
\end{enumerate}

In principle the signal significance can then be calculated either with forward-modeling likelihood fits or with Bayesian sampling. When characterizing signal significance, however, it is common to make a number
of simplifying assumptions:

\begin{itemize}
\item Fixing the instrument response, i.e., not fitting for parameters in
  the instrument response;
\item Fixing all source fluxes except a test source and a large-scale
  diffuse background;
\item Fixing the spectrum of the test-source, e.g., to be a
  fixed-index power-law.
\end{itemize}

With these simplifying assumptions, it is possible to model a region of
interest with a single free parameter for the source (i.e., the
normalization), and one parameter for the background.  (i.e., the normalization
of the large scale background).

This makes it tractable to analytically compute good approximations to the likelihood
function, and often to find analytic forms for the derivatives of the
likelihood.   In some cases it is even possible to find an analytic expression
or semi-analytic expression for the set of parameters that maximize the
likelihood, and estimate curvature of the likelihood (and hence the
parameter uncertainties) near the maximum.   

For an example of such a computation, see the ``semi-analytic
sensitivity estimate'' provided in Ref.~\cite{2FGL}.

In spirit, these methods are extensions of photon counting methods, i.e., estimating the
 number of background events that lie in the signal region, and then estimating the
signal significance based on any excess of events in the signal region with respect to that
background estimate.

\subsection{Simulating events using IRFs}

If the IRFs are tabulated, it is often much faster to simulate
events using the tabulated IRFs than a full simulation of the
instrument.   If, for example, the IRFs can be factored in separate
pieces representing the effective collecting area ($A_{\rm eff}$), the
point-spread function ($P(\hat{\nu}'|\hat{\nu})$) (i.e., the probability to assign a
direction $\hat{\nu}'$ given the $\hat{\nu}$)  and the energy
dispersion $D(E' | E) $ (i.e., the probability to
reconstruct the event energy as $E'$ given the true energy
$E$), then simulating events can be factored into generating events with true energy and directions $E, \hat{\nu}$
given the source characteristics and the $A_{\rm eff}$, and then use the PSF and the energy dispersion
to smear $E, \hat{\nu}$ into  $E', \hat{\nu}'$.

Unfortunately, the extra dimensionality of the Compton data space used to analyze Compton scattering 
events makes it impractically difficult to generate and use IRF tables in the same way
for Compton data analysis.

\subsection{Simulating maps using IRFs and exposure tables}

If the ``observing profile'' (the amount of time a given part of the sky 
was observed at a particular angle in the instrument reference frame) or the 
``observing history'' (the amount of time a source was observed under particular
observing conditions) is known, and
corresponding IRFs for all of the observing conditions are also known, then,
at the cost of the loss of some information, fast simulations can be
produced by first creating ``mean'' IRFs; in other words, IRFs that are averaged over the observing profile or observing history. These IRFs can then be applied to binned maps of the sky.

This technique is used extensively in the {\it Fermi}-LAT data analysis and particular in the 
Fermipy software suite\footnote{https://fermipy.readthedocs.io/en/latest/}.    One of the standard data products generated by the
mission provided Fermitools\footnote{https://fermi.gsfc.nasa.gov/ssc/data/analysis/} 
is the ``livetime cube", which is actually a table that gives the amount of time each direction ($\alpha, \delta$) in the 
sky was observed at a particular angle $(\theta)$ with respect to the LAT boresight: $t_{\rm live}(\theta 
| \alpha, \delta)$.   Other standard tools generate the mean IRFs and convolve them with 
flux models of the sources being analyzed to generate templates that can be used to compute
the maps of the number of photons observed for a given source flux.  Particular points to note
here are that by using a single, averaged, set of IRFs for the entire observation, the smearing can be simplified
to simply convolving the maps with the PSF; and similarly, the energy dispersion can be reduced to
applying at energy redistribution matrix to the energy counts per bin.   This avoids having to go back to the individual
events, and greatly speeds that data analysis.

It is also very important to understand that maps generated using these fast-simulation techniques
are extremely useful, as they provide an expected number of photons per pixel and per energy, 
i.e. $n_{\rm expect}(E, \alpha, \delta | {\rm flux, model})$, and a great many simulated observations can be generated simply
by throwing Poisson-distributed counts with respect to $n_{\rm expect}$, which can be critical for computing expectation bands
or doing Bayesian parameter estimation.

\section{Simulation verification and limitations}

Simulations can never perfectly replicate real data. Air shower physics are complex and the models used for simulations all make various simplifying assumptions. Models of the atmosphere can at best approximate the actual atmospheric profile, and detector models can never hope to reproduce all the small defects, mis-alignments, inefficiencies etc. of a real detector. Yet, even with these limitations, simulations are still a very useful tool, as long as we acknowledge and study their limitations.

Predictions from simulations are typically compared to measurements to validate the simulations. The validation procedure can include calibration measurements in the lab, at testbeams, or in flight/in situ, as well as comparisons of certain distributions such as gamma-hadron separation parameters and the measured direction of reconstructed photons from known point-like gamma-ray sources. Measured fluxes and energy spectra can be compared to results from other instruments (cross-calibration). Often, dedicated simulations of for example a test beam setup are needed.  The detector model and detector effects engine used in simulations typically undergo multiple revisions and and are tweaked until they match the measurements. Parameters of the detector model that can not be fully constrained by calibrations and other measurements should be considered as part of the systematic uncertainty of the instrument. This can be done for example by preparing alternative sets of instrument response files based on simulations with certain parameters in the detector model changed with respect to the nominal. The Crab nebula, one of the strongest known sustained gamma-ray emitters, is often used to validate the final instrument model, see for example \cite{2016_magic2, MAGIC_LZA, HAWC_2017, Abeysekara_2019, Meagher:2015igh, VTS_throughput_2021, HESS:2006fka, HESS:2015cyv}. Other strong sources such as gamma-ray pulsars and active galactic nuclei can be used as well \cite{2012ApJS..203....4A}. 

Background simulations can be validated against measurements on source-free regions on the sky (or source-free time intervals). However, backgrounds from cosmic rays and detector noise are notoriously hard to predict well in simulations. It is thus preferred to use measured background distributions whenever possible (after launch/inauguration). 


\section{Summary}

Many different types of simulations are crucial pieces of the tool kit for high-energy astronomy.   They are used to develop the case for new instruments, to design those instruments, to develop the data reduction and event analysis pipelines for those instruments, to construct the IRFs used to analyze the data from those instruments, and to generate the simulated datasets needed to perform high-level analysis. Because of the ubiquity of the need for simulations coupled with the variety of the particular details relevant for any given type of instrument, as well as the variety of the different contexts in which simulations are used, a great many simulation tools exist within high-energy astronomy. In this chapter, we have described the common aspects of simulations, given a short survey of the available simulation tools, and discussed how simulations are used for instrument design and data analysis in high-energy astronomy.

\section{Cross-References}

Orbits and background of gamma-ray space instruments\\
Compton telescopes\\
Pair-creation telescopes\\
The COMPTEL instrument on the CGRO mission\\
The AGILE mission\\
The Fermi Gamma-ray Space Telescope mission\\
The Large Area Telescope on the Fermi mission\\
HAWC: The High Altitude Water Cherenkov detector\\
The Major Gamma-ray Imaging Cherenkov Telescopes (MAGIC)\\
 The Very Energetic Radiation Imaging Telescope Array System (VERITAS)\\
 The High Energy Stereoscopic System (H.E.S.S.)\\
 The Cherenkov Telescope array (CTA): a worldwide endeavor for the next level of ground-based gamma-ray astronomy\\
 
\section{Acknowledgments}
C. S. acknowledges support by the NRL Isabella and Jerome Karle Distinguished Scholar Fellowship Program. H. F. acknowledges support by NASA under award number 80GSFC21M0002.



\renewcommand{\refname}{Citations}

\bibliography{main}

\end{document}